\documentclass[aps,preprint,amsmath,amssymb,amsfonts]{revtex4}
\usepackage{epsfig}
\usepackage{graphicx}
\usepackage{subfigure}
\usepackage{dcolumn}
\usepackage{bm}
\usepackage{amsthm}
\usepackage{enumitem}
\usepackage{slashed}
\usepackage{braket}

\newcommand{\bbR}{\mathbb{R}}

\begin{document}

\title{Fine-tuning and the Second Law of Thermodynamics from the perspective of de Sitter quantum gravity}

\author{Yasha Neiman}
\email{yashula@gmail.com}

\affiliation{Okinawa Institute of Science and Technology, 1919-1 Tancha, Onna-son, Okinawa 904-0495, Japan}

\date{\today}

\begin{abstract}
In this essay, we discuss the fine-tuning problems of the Higgs mass and the cosmological constant. We argue that these are indeed legitimate problems, as opposed to some other ``problems'' that are sometimes described using similar vocabulary. We then notice, following Tom Banks, that the problems become less compelling once we recognize that the Universe contains quantum gravity, and thus isn't fundamentally described by bulk QFT. Embracing this ``solution'' requires a reversal of the standard arrows UV$\rightarrow$IR and past$\rightarrow$future. The first reversal is familiar from AdS/CFT. The second reversal refers more specifically to our Universe's cosmology, and is clearly in potential conflict with the Second Law of Thermodynamics. In the final part of the essay, we attempt to defuse this conflict, suggesting that the Second Law can arise naturally from de Sitter boundary conditions at future infinity.
\end{abstract}

\maketitle

\section{Introduction and summary} \label{sec:intro}

This is an essay rather than a research paper. Rather than present new facts, I would like to offer a perspective on what's already known. A great many theoretical efforts have been expended on explaining, via various mechanisms, the smallness (in Planck units) of the Higgs mass $m_H$ and of the cosmological constant $\Lambda$. Indeed, the various proposed mechanisms for ``justifying'' a light Higgs, such as weak-scale supersymmetry, arguably make up the majority of beyond-Standard-Model predictions that are being tested at the LHC. As for the smallness of the cosmological constant, it is widely presented in textbooks as ``the greatest discrepancy in physics''. 

In this essay, I will argue that, once quantum gravity and holography are taken into account, these concerns begin to seem misguided. There is nothing new about this statement -- for example, see \cite{Banks:2000fe}, as well as the recent \cite{Banks:2018jqo}. And yet, it is a rare statement to hear, making it worthwhile to spell it out. The key point is that quantum gravity subverts the typical roles of short distances (UV) and long distances (IR), allowing us to view the small IR values of $m_H$ and $\Lambda$ simply as fundamental parameters of the theory. In this way, the fine-tuning problems can be ``delegated'' to the realm of quantum gravity and holography. This clearly isn't a constructive solution, since quantum gravity in the real world is an enormous open problem in its own right. However, it's a problem that will need to be solved \emph{anyway}. What the argument of e.g. \cite{Banks:2000fe,Banks:2018jqo} is saying is that fine-tuning isn't an \emph{additional, separate} problem. Thus, it need not have a separate solution, either within classical gravity (as in long-range modifications of GR motivated by the fine-tuning of $\Lambda$), or within non-gravitational QFT (as in extensions of the Standard Model motivated by the fine-tuning of $m_H$). Therefore, if we agree with this argument, we must significantly update our biases vis. the theoretical necessity of such solutions, as well as our expectations vis. their experimental confirmation.

Now, how would embracing this ``solution'' to the fine-tuning problems affect the rest of our world-picture? The reversal of roles between UV dynamics, normally thought of as fundamental, and IR dynamics, normally thought of as emergent, is a familiar basic feature of AdS/CFT \cite{Witten:1998qj,Aharony:1999ti}. Indeed, if we lived in anti-de Sitter space (AdS), such a reversal would be quite uncontroversial. On the other hand, within \emph{our} Universe's cosmology, interchanging UV and IR carries deeper consequences. With a UV-dominated Big Bang in the past and an IR-dominated de Sitter future, reversing the UV$\rightarrow$IR arrow implies also reversing the roles of initial and final conditions, i.e. \emph{reversing the arrow of time}. It is important that the \emph{only} potential conflict between such a reversal and any known fundamental physics is the Second Law of Thermodynamics. But, of course, such a conflict must be taken seriously, and at first sight it may appear fatal to the entire ``backwards'' worldview of \cite{Banks:2000fe,Banks:2018jqo}. 

Note that the challenge from the Second Law is not just to quantum gravity / holography \emph{as a solution to the fine-tuning problems}. It is a challenge to the entire broad notion of describing our world holographically, since the most plausible foothold for such a description is at the future boundary of our asymptotically de Sitter spacetime. In the essay's final part, we will argue that this challenge can in fact be met. In other words, we will point out a worldview in which the thermal arrow of time arises naturally from an opposite fundamental arrow. Our picture will depend on some separation and decoupling of scales, which we know to be true properties of the real world. We do not claim to quantitatively explain the ``strength'' of the Second Law, i.e. the precise degree of smallness of the early Universe's entropy. Our goal is simply to point out that a low-entropy past, with entropy increase towards the future, can arise from de Sitter asymptotic conditions in the infinite future.

As the reader has already intuited, we will assume that our Universe is indeed future-asymptotically de Sitter, i.e. that the simplest interpretation of its observed accelerated expansion \cite{Riess:1998cb,Perlmutter:1998np} is correct. A further assumption will be that something like a dS/CFT correspondence \cite{Strominger:2001pn} holds. We not pretend to know how (or whether) dS/CFT should work in the real world. We simply take it as the most straightforward combination of what is known theoretically about quantum gravity, and what is known observationally about cosmology. 

While the assumption of a future-asymptotic de Sitter is utterly straightforward from the observational point of view, it is highly loaded and contested in the theory community. There exists various indirect evidence for a conflict between de Sitter space and string theory (see e.g. \cite{Danielsson:2018ztv}), and even between de Sitter gravity and quantum mechanics itself (e.g. quantum mechanics requires an observer with asymptotically large memory, which isn't available inside a de Sitter horizon). We will address this matter further in the Discussion section. For now, we just assume a de Sitter future, trusting that the rest of physics will adapt as it must. If this assumption is wrong, then some of the arguments below will still be relevant, but the overall worldview that we'll suggest is probably doomed.

When referring to de Sitter space, we will always mean the observed present \& future de Sitter phase of the Universe, as opposed to the conjectured inflationary one. As far as we can tell, our arguments are independent on whether or not inflation occurred. Our logic \emph{will} run strongly against using flatness or homogeneity as arguments in favor of inflation. However, there are better reasons to believe in inflation, and we will not consider this issue further.

The rest of the essay is structured as follows. In section \ref{sec:tuning_vs_hierarchy}, we clear some fog regarding what is a legitimate fine-tuning problem. In section \ref{sec:arrows}, we focus on the assumptions that give the fine-tuning problems their teeth; these are the directionality of the UV$\rightarrow$IR renormalization flow and of past$\rightarrow$future evolution. We then discuss how both assumptions can be undermined, potentially dispelling the fine-tuning problems. In section \ref{sec:second_law}, we argue that a Second Law of Thermodynamics can paradoxically arise from a future$\rightarrow$past fundamental worldview. In section \ref{sec:discuss}, we discuss some of the many open ends.

\section{What makes a fine-tuning problem?} \label{sec:tuning_vs_hierarchy}

Many beginning students, as well as established physicists in other fields, are puzzled by the fundamental theorists' obsession with fine-tuning problems. In part, this attitude may be attributed to a genuine difference in values. However, I believe that much of it is due to at unfortunate conflation between legitimate causes for concern and similar-sounding illegitimate ones. The issue is related, though imperfectly, to theorists' usage of the alternative terms ``hierarchy problem'' and ``naturalness'', which lend themselves more easily to misinterpretation than ``fine-tuning''.

It's all too typical for somebody -- let's call them Physicist A -- to assert that the \emph{very fact} that $m_H$ and $\Lambda$ are much smaller than the Planck scale is a ``hierarchy problem'', because numbers much smaller than 1 are ``unnatural'', without any further explanation. Physicist B will then respond that no, we cannot know whether a small number is natural or not, since that depends on an assumed flat probability distribution over the possible values. We do not know the actual distribution, B will argue, or even whether there's a distribution to discuss at all. When the debate is phrased like this, I am wholeheartedly on B's side. Large or small numbers \emph{as such} are not a problem. A theory is allowed to have a very small (or very large) number as a parameter. If we discover suddenly that Nature is described e.g. by a gauge theory with a very large number of colors $N$, or with a very small or very large coupling $g$, this \emph{in itself} should not be cause for drama. To make a motto of it: hierarchies are facts, not problems.

Of course, the real reason why we're worried about the smallness of $m_H$ and $\Lambda$ is more sophisticated: it is that their small values are \emph{not} fundamental parameters, but rather \emph{extremely unlikely emergent outcomes} of renormalization flow. Specifically, they must have large ``bare'' values in the UV (actually, order-1 in Planck units), tuned to great precision \emph{so as to produce}, upon coarse-graining, the small observed values in the IR. When stated in these more specific terms, the smallness of $m_H$ and $\Lambda$ \emph{is} a valid reason to lose sleep. It is this problem -- the necessity of fine-tuning, the miraculous cancellation of large numbers to produce small numbers, rather than the mere \emph{existence} of small numbers -- that we must resolve.

More generally, a true fine-tuning problem can be viewed as a failure of the scientific reductionist worldview. Science is about finding simple explanations for Nature's (usually complicated) emergent phenomena. This program can conceivably fail, and if it seems to do so, we must indeed sound the alarm and begin scrambling for solutions. In order to declare that we are faced with such a failure, the following conditions must be satisfied:
\begin{enumerate}
	\item The emergent phenomena have some striking, simply stated property. In our case, these are the small IR values of $m_H$ and $\Lambda$.
	\item This striking property is unstable under slight perturbations at the more fundamental explanatory level. In our case, a slight change in the UV values of $m_H$ and $\Lambda$ would spoil their smallness in the IR.
	\item The unlikely setup that enables our emergent property \emph{does not appear special at the fundamental level}, i.e. it has no obvious description without referring to the emergent property itself. In our case, the necessary UV values just look like random order-1 numbers, specified to a ludicrous level of accuracy.
\end{enumerate}
When understood in this way, the concept of fine-tuning problems can be usefully extended beyond quantum field theory, and even beyond the quantitative sciences.

\subsection{Life: a very different fine-tuning problem}

Let us then ask: have we, in the history of science, ever encountered a fine-tuning problem, other than those of $m_H$ and $\Lambda$? I am aware of one example -- the existence of living things. Before Darwin and Wallace, the world of living organisms satisfied all of our criteria for a fine-tuning problem. A living creature has some striking properties, with simple emergent descriptions such as ``good at running'', ``good at jousting'', or even ``good at storytelling''. On the other hand, an organism is made of atoms, following the laws of physics from some initial conditions. The organism's special skills require some very non-generic configurations of these atoms: the vast majority of atomic configurations aren't good at anything, i.e. they are dead. And yet, we can't describe \emph{in fundamental terms} what it is that makes a living heap of atoms special -- it can only be described via emergent concepts such as ``running''. This fine-tuning problem was quite real and urgent, until its resolution in the 19th century via the insight of natural selection.

The case of living things can be rephrased in other terms, which will be useful below. Instead of fundamental and emergent levels of explanation, we can speak of causes and effects. Living things behave as if they have \emph{purposes}, which makes them seem out of place in a world of \emph{causes}. This illustrates a general lesson: a fine-tuning problem can be thought of as an \emph{apparent reversal of causality}. The achievement of Darwin and Wallace was to show how the appearance of purpose can in fact arise from past causes -- in this case, by shifting the focus from the organism to its ancestors.

\subsection{Non-problems: initial flatness, homogeneity and the Second Law}

Let us now discuss some situations in physics that are sometimes advertised as fine-tuning problems, but, in our opinion, do not survive close scrutiny. One cluster of such ``problems'' are the near-perfect spatial flatness, homogeneity and isotropy of the Big Bang. These are striking properties of our Universe, which are very non-generic, in the sense that they'd be violated by most configurations of gravity and matter. However, they can be phrased directly as properties \emph{of the Big Bang's initial conditions}, with no need to refer to any subsequent, emergent phenomena. A possible source of confusion is that these properties of the Big Bang were in fact \emph{necessary} -- at least to some degree -- for the emergence of the large and complex Universe we know today. However, there's no need to say, in violation of reductionism, that the Big Bang was ``fine-tuned so as to enable the emergence of rabbits''. It is sufficient to make the simple, fundamental-level statement, ``the Big Bang was very flat, homogeneous and isotropic''. The far-reaching consequences of this simple initial property are not a problem, but a classic \emph{success} of scientific reductionism! In other words, the Big Bang's remarkable flatness and homogeneity are ``hierarchy facts'', not fine-tuning problems.  

A similar discussion applies to the Second Law of Thermodynamics. The Universe's small initial entropy is often touted as a mystery requiring explanation. Until recently, I myself was guilty of this. Of course, it would be wonderful to find a mechanism behind this low entropy, and our section \ref{sec:second_law} might qualify as an attempt. And yet, the lack of such a mechanism is \emph{not} a gaping incompleteness in our understanding of Nature. Unlike dark matter or quantum gravity, it doesn't point to any inconsistency within our existing knowledge. Yes, a low-entropy initial state is (by definition) very non-generic. But nothing about it defies reductionism. The initial conditions' low entropy can be cleanly stated in terms of the initial conditions \emph{themselves}, with no need to invoke the complex emergent structures that are subsequently enabled by the Second Law.

In fact, in cosmology as we know it, the Second Law is not merely \emph{analogous} to the initial flatness, homogeneity and isotropy: it's largely \emph{equivalent} to them. It is the homogeneity of the Big Bang that accounts for its low entropy, which then increases in the process of gravitational clumping, followed by star formation etc. This is of course well-known, but isn't celebrated as much as it should be. Before the first observations of the Cosmic Microwave Background, the precise nature of our Universe's low initial entropy was mysterious. But now that we observed the CMB and found it to be homogeneous, the nature of the Big Bang's low entropy is known! It is also worth noting that such a simple, elegant form of the low-entropy initial state is only possible thanks to gravity. Without gravity, a homogeneous state would be the one with \emph{maximal} entropy. In a non-gravitating Universe, low-entropy initial states actually sound rather contrived. Should it be some gas with all atoms arranged in a regular lattice? Or a soup of unstable nuclei, which then start to decay? In contrast, gravity provides us with a beautiful -- and observationally confirmed -- low-entropy initial state: a smooth \& homogeneous Universe.

\section{Reversing the two arrows: (UV$\rightarrow$IR) and (past$\rightarrow$future)} \label{sec:arrows}

\subsection{The (UV$\rightarrow$IR) arrow}

We now return to the fine-tuning problems of $m_H$ and $\Lambda$. Let us list their ingredients a bit more explicitly. 
\begin{enumerate}
	\item \emph{Observational fact:} the IR values of $m_H$ and $\Lambda$ are small.
	\item \emph{Theoretical fact:} under renormalization flow, $m_H$ and $\Lambda$ change by a large amount.
	\item \emph{Inference from 1 and 2:} the UV values of $m_H$ and $\Lambda$ are large.
\end{enumerate} 
By themselves, these facts don't yet amount to a fine-tuning problem. The problem appears when we add one final ingredient:
\begin{enumerate}
	\item[4.] \emph{Assumption:} the UV values are fundamental, while the IR values are emergent.
\end{enumerate}
With this assumption, we obtain the fine-tuning problem:
\begin{align}
 (\text{large UV value}) - (\text{large RG flow}) = (\text{small IR value}) \ . \label{eq:tuning}
\end{align}
On the other hand, if we start with the reverse assumption:
\begin{enumerate}
	\item[4'.] \emph{Assumption:} the IR values are fundamental, while the UV values are emergent,
\end{enumerate}
then eq. \eqref{eq:tuning} rearranges into:
\begin{align}
 (\text{small IR value}) + (\text{large RG flow}) = (\text{large UV value}) \ . \label{eq:no_tuning}
\end{align}
There is no fine-tuning here! In eq. \eqref{eq:no_tuning}, we simply have an initial small value, which undergoes a violent process (i.e. the RG flow), and becomes large. One might think of eq. \eqref{eq:no_tuning} as an egg breaking, while eq. \eqref{eq:tuning} is like an egg putting itself together. From this point of view, a fine-tuning problem is just a movie played backwards -- a reversal of causality, as we noted in the previous section. If we could just flip the direction in which we think about RG flow, and set the IR parameters as fundamental, the problem would completely disappear.

The issue, of course, is that such a reversal runs against established knowledge of QFT. In QFT, setting the UV dynamics as fundamental is not an arbitrary choice: the RG flow \emph{must} run from UV to IR (barring perhaps some special cases \cite{Cavaglia:2016oda}). The crucial counterpoint to this, which for some reason is often unacknowledged, is that the physics of our world is ultimately \emph{not a QFT}, or at least not one that lives in the 3+1d spacetime of everyday experience. The reason is that our world includes \emph{gravity}. In quantum gravity, a very general wisdom is that the clean QFT notions of UV and IR do not survive. A standard simple way to see this is to notice that the Schwarzschild radius grows with mass, unlike the Compton wavelength, which decreases. 

The statement is even sharper in the best-understood working models of quantum gravity, namely in AdS/CFT. There, the quantum-gravitating universe \emph{is} described by a QFT, but it is a lower-dimensional QFT that lives on the spacetime's conformal boundary. Within this boundary conformal field theory (CFT), the standard field-theoretic wisdom applies: renormalization runs from UV to IR. However, crucially, the holographic dictionary \emph{inverts} the UV/IR labels: what appears as UV on the boundary is IR in the bulk. Thus, in the bulk, we shouldn't simply \emph{entertain the notion} of IR parameters being fundamental; we should positively \emph{expect} this to be the case. And, indeed, this is how AdS/CFT operates: from the CFT, we can read off directly the $n$-point correlators of bulk fields \emph{as measured at large distances}. We are never forced to think of them as emerging from short-distance bulk physics. In fact, this short-distance physics -- presumably a complete bulk formulation of string theory -- remains largely unknown. 

The most striking and relevant example of this general principle is the AdS cosmological constant. In AdS/CFT, its value in Planck units is essentially just a negative power $|\Lambda|\sim N^{-\alpha}$ of the number of colors $N$ in the CFT. In other words, the \emph{IR value} of the cosmological constant is \emph{directly determined} by a fundamental parameter of the boundary theory! Small $|\Lambda|$ simply means large $N$. And, in the terminology of section \ref{sec:tuning_vs_hierarchy}, large $N$ is \emph{not} a fine-tuning problem: it's just a hierarchy fact.

To summarize, quantum gravity in general undermines Assumption 4, while AdS/CFT specifically realizes the opposite Assumption 4'. In a universe described by something like a boundary CFT, bulk IR parameters become fundamental. In such a universe, $m_H$ and $\Lambda$ wouldn't be fine-tuned -- they would instead be fundamental parameters, which simply happen to be small.

\subsection{The (past$\rightarrow$future) arrow}

A cosmologist might react to the previous subsection with impatience. For them, the running of e.g. the cosmological constant is not some abstract issue of renormalization flow, but an actual process that occurs through cosmological history. For instance, we're quite sure that, in the early Universe, there occurred an electroweak phase transition. In that transition, the Higgs field acquired its nonzero vacuum expectation value, which shifted the vacuum energy density by an amount of the order $(100\text{GeV})^4$. We must then conclude that \emph{before} the transition, the vacuum energy density (i.e. the cosmological constant) had a very fine-tuned value of that order of magnitude, which canceled near-perfectly with the effect of the phase transition (and any subsequent contribution from a QCD phase transition), to yield the tiny value observed today. Thus, the fine-tuning problem arises not just when we coarse-grain from UV to IR, but also when we simply follow cosmic history from past to future.

On the other hand, if we reverse also \emph{this} arrow, i.e. evolve the Universe from future to past, then the fine-tuning again disappears. In such a reversed view of history, the Universe starts out with small effective values of $m_H$ and $\Lambda$ in the future, then undergoes some violent phase transitions, which yield large values in the past. Eq. \eqref{eq:tuning} is replaced once again by eq. \eqref{eq:no_tuning}, and all is well. 

But does it \emph{make sense} to reverse the arrow of time in this way? Does it make sense to work with final, rather than initial, conditions? From the point of view of all known dynamical laws, the answer is absolutely yes. While reversing the UV$\rightarrow$IR arrow of QFT required a recourse to quantum gravity, there is \emph{nothing whatsoever} in either General Relativity or the Standard Model that favors the past$\rightarrow$future arrow over its opposite. The reason we think in terms of this arrow at all is the Second Law of Thermodynamics: entropy is smaller in the past. In particular, the fact that our minds (along with their environment) are subject to the Second Law is the reason why we can't remember the future. At this juncture, some might claim that \emph{quantum measurement} is another phenomenon that prefers initial over final conditions. My point of view is that the non-reversibility of quantum measurement is just the non-reversibility of decoherence, which in turn is just another manifestation of the Second Law. In the low-entropy past, the Universe has non-entangled systems; as we evolve towards the future, these gradually become entangled, which is the generic, high-entropy situation.

To sum up, we can reverse the past$\rightarrow$future arrow with a clear conscience, as long as we ignore the Second Law. We will now briefly do just that, and examine the resulting worldview. We will pick up the Second Law once more in section \ref{sec:second_law}.

\subsection{Both arrows reversed in dS/CFT}

We've seen that the known facts about $m_H$ and $\Lambda$ take the shape of a fine-tuning problem \emph{if} we assume that UV couplings and initial conditions are fundamental, with the IR couplings and final conditions emerging dynamically. If we reverse those arrows, the fine-tuning problems evaporate. In our Universe's cosmology, the two arrows are actually tied  together: in the past, there is a Big Bang, where UV dynamics dominates; in the future, there is asymptotic de Sitter space, where IR dynamics dominates. Thus, if we reverse one arrow, we should also reverse the other. For the past$\rightarrow$future arrow, fundamental physics (excluding for now the Second Law) gives us \emph{no} indication for its correct direction, so we can simply reverse it. For the UV$\rightarrow$IR arrow, we \emph{do} have a direction dictated by QFT; however, that clear statement is rendered fuzzy by quantum gravity, and gets completely turned on its head by holography.

Until now, when invoking holography, we referred to AdS/CFT -- the one case that's truly under theoretical control. Let us now make the (not unreasonable?) leap of faith that something much like AdS/CFT holds also for our Universe. This would mean that our world is described by dS/CFT, where the conformal boundary is no longer at AdS \emph{spatial} infinity, but at the \emph{future} infinity of our asymptotically de Sitter spacetime. Once we accept this picture, both of our arrows are naturally reversed! The Universe is now defined fundamentally by some boundary theory at future infinity, where all bulk distances are large. From there, we have a notion of evolution into the past, analogous to inwards radial evolution in AdS/CFT. In particular, we expect that the bulk evolution from the IR future into the UV past should correspond to some (standard, non-reversed) UV$\rightarrow$IR coarse-graining flow in the boundary CFT. Eventually, this evolution takes dramatic turns, leading to various phase transitions (such as the electroweak one), and ultimately to a Big Bang. As a result of such violent processes, the fundamental small values of $m_H$ and $\Lambda$ become order-1. The fundamental small values themselves are just reflections of some very large (or small) dimensionless parameters in the boundary theory, such as a large number of colors $N$ in the case of the cosmological constant. And, again, there's nothing wrong with extreme values for fundamental parameters -- they are just ``hierarchy facts''. The story is free of fine-tuning.

\section{A Second Law after all} \label{sec:second_law}

As we've seen, in AdS/CFT, the fundamentally correct direction of RG flow (speaking in terms of bulk scales) is IR$\rightarrow$UV. However, the naive UV$\rightarrow$IR flow is still a valid property of a certain approximate, effective description, namely of bulk field theory (which, for the present discussion, is meant to include the gravitational sector). For an AdS bulk observer who only probes length scales much smaller than the AdS curvature radius and much larger than the string \& Planck lengths, it is the UV$\rightarrow$IR flow that makes more immediate sense, even though, fundamentally speaking, it's ``backwards''. The precise mechanism by which this ``backwards flow'' emerges at intermediate scales is still mysterious to some extent (and is essentially equivalent to the emergence of bulk locality). And yet, if we believe all the existing evidence for AdS/CFT, we must conclude that such a backwards flow indeed emerges.

Can a similar ``miracle'' happen for the thermal arrow of time? Can a past$\rightarrow$future Second Law emerge, as an approximate phenomenon at intermediate scales, out of a fundamental future$\rightarrow$past flow? As we will now argue, the answer seems to be yes.

To begin, let us recall a very basic, well-known circumstance. The low-curvature de Sitter spacetime which we observe at present (and extrapolate into the future) can only arise from some very special initial conditions at the Big Bang. In particular, the Big Bang must be very smooth and homogeneous, so as to prevent violent curvature fluctuations. Even the spatial flatness of the Big Bang (and thus of our entire FRW cosmology) can be related to a property of the de Sitter future, namely an $\bbR^3$ topology of the conformal future boundary. This dependence of the de Sitter future on special initial conditions is often presented as a fine-tuning problem. For reasons already spelled out, we reject that characterization. However, this dependence is still important. In particular, the special initial conditions which enable the de Sitter future are \emph{qualitatively the same} as the special initial conditions that constitute our Universe's low initial entropy! While this observation isn't new, it is worth spelling out its implications for a worldview with a fundamental future$\rightarrow$past arrow of time.

Once again, our fundamental assumption is a de Sitter asymptotic future, with the currently observed small cosmological constant. Moreover, we assume that this Universe is somehow defined by a CFT at future infinity. Coarse-graining in the CFT generates evolution into the bulk, i.e. into the past. Eventually in the course of this evolution, we encounter a Big Bang. As long as we stay away from the past singularity, the Big Bang is well-described by local bulk physics. However, this bulk physics isn't fundamental: it is merely an effective description, ultimately arising from the fundamental theory at future infinity. In this picture, the dependence between the de Sitter future and special initial conditions is still present, but in reverse: instead of the de Sitter future \emph{requiring} a very special Big Bang, it \emph{induces} a very special Big Bang. More precisely, it induces a Big Bang that is restricted to a tiny portion of the phase space \emph{of the effective bulk theory}. In other words, the effective bulk theory severely overestimates the range of possible states: most of the states that seem legitimate in the effective description can never arise from the fundamental one. This situation is similar to the ``swampland'' conjectures in string theory \cite{ArkaniHamed:2006dz}, according to which most low-energy QFT's are actually ruled out by the requirement of an underlying string theory (or, perhaps more generally, an underlying quantum gravity). However, here we are concerned not with the choice of the effective theory itself, but rather with the choice of state within it. 

The final ingredient for a bulk Second Law will be a separation of scales. As our Universe expands and cools, the effective degrees of freedom become less and less correlated with the de Sitter future. In the early Universe, even small deviations from homogeneity could spoil the low-curvature future. However, the de Sitter future is not so sensitive to the behavior of matter at later stages, i.e. at intermediate scales. Dust, stars, galaxies, organisms, civilizations -- these can all form, merge and be destroyed, without affecting the de Sitter future. As a result of this decoupling, as the Universe expands, more and more of the effective theory's phase space at the relevant scales is ``opened up'', in the sense that it's no longer ruled out by the de Sitter future.

This completes all the necessary ingredients for a Second Law of Thermodynamics. In the past, our future boundary conditions impose a special, low-entropy state. As we progress into the future, the restriction is loosened, so the effective bulk theory is freed to explore its phase space, and increase its entropy. In this way, a Second Law running from past to future can arise naturally from a Universe that's fundamentally defined by a CFT at future infinity.

The argument above is clearly qualitative. In particular, we do not claim that the low entropy of the real world's Big Bang was \emph{as large as it could be} under the constraints of the de Sitter future. We simply aimed to show that an initially low and subsequently increasing entropy can, \emph{as a matter of principle}, arise from a reversed fundamental arrow of time.

Before closing this section, let's address one common objection. The asymptotic future in de Sitter space is often presented as a \emph{high}-entropy state; in that case, how can it induce a special, low-entropy state in the past? This seems to be a confusion that disappears upon closer inspection. Indeed, the asymptotic future has high entropy \emph{within the Hilbert space of a de Sitter causal patch}. This doesn't contradict the statement that having a low-curvature de Sitter future \emph{at all} is a very non-generic outcome, at least from the perspective of the effective bulk theory in the early Universe. Note also that our argument does not require the effective theory's entropy counting to be \emph{fundamentally correct}. In fact, we expect the holographic de Sitter theory to be completely self-contained, so that within it, the de Sitter future is an inviolable fact, and states that are incompatible with this future are ``not really there''. For the Second Law to emerge via the mechanism we're suggesting, it is enough that it just \emph{seems}, from the bulk theory's vantage point, that the states which end up leading to de Sitter asymptotics are a tiny minority.

\section{Discussion} \label{sec:discuss}

In this essay, we argued that a perceived crisis in fundamental theory -- the fine-tuning problems of $m_H$ and $\Lambda$ -- may be illusory. We then probed the worldview that arises if one takes this possibility seriously. We concluded that it leads to an unusual -- but perhaps consistent -- viewpoint on the ladder of scales and the arrow of time. 

One can imagine physical processes that would invalidate the worldview advocated here. For instance, we assumed that the Universe won't undergo any further phase transitions in the future. Generically, such a transition would carry $\Lambda$ away from its present small value. For example, this can occur if some (natural or artificial) event triggers a second electroweak phase transition, as was suggested e.g. in \cite{Masina:2012tz}. I can wrap my head around a small \emph{final} value of $\Lambda$, but not around a small value that just happens to occur \emph{in the middle} of cosmic history. If the latter turns out to be the case, then we are back in fine-tuned territory.

We must also acknowledge a certain elephant in the room: a sizable proportion of theorists \emph{expect} our Universe to eventually exit the present de Sitter phase -- not because of some observational indication, but because a quantum-gravitational de Sitter space is viewed as a \emph{theoretical impossibility} -- see e.g. \cite{Obied:2018sgi}. There are at least two kinds of argument in this direction. The first is specific to string theory, and concerns the difficulty of constructing de Sitter string vacua (with the status of the KKLT proposal \cite{Kachru:2003aw} still unclear). The second kind does not refer to string theory, suggesting more directly a conflict between de Sitter gravity and quantum mechanics itself. One argument in this vein is that, within relativistic physics, quantum commutators are intimately tied to spacetime's causal structure, which, in quantum gravity, itself becomes uncertain. This can be overcome in asymptotically flat or AdS spacetimes, where a classical causal structure exists at conformal infinity. In contrast, in asymptotically de Sitter space, conformal infinity is spacelike, and causal structure only exists in the bulk, where the geometry is subject to quantum uncertainty. This suggests that de Sitter may be incompatible with (exact) quantum mechanics. Another such argument, often made by Nima Arkani-Hamed, concerns the finite dimension of the Hilbert space that is available to a de Sitter observer: within such a Hilbert space, there is no room for an idealized ``classical measurement apparatus'' with infinite memory, so that one can never test \emph{with asymptotic precision} any prediction of quantum physics. 

If the above arguments are true, \emph{and} one believes in the sanctity of quantum mechanics (or of string theory), then one must disbelieve the asymptotic de Sitter future, and the portions of this essay that rely on it become irrelevant. My personal bias is different. I believe we should accept the observational evidence of a positive cosmological constant, and entertain the possibility that quantum mechanics is not exact. Of course, this viewpoint does not imply any significant deviations from QM at accessible scales. For instance, if the ultimate failure of QM is due to the finite-dimensional Hilbert space of a de Sitter observer, then we should not expect to notice its effects much earlier than the de Sitter recurrence time, which is exponentially long (see e.g. \cite{Dyson:2002pf}).

This last point relates to a general source of unease about the picture proposed here. In this essay, we referred to the infinite future of de Sitter space, as well as to our present and past eras, without really acknowledging the vastness of time between now and future infinity. The de Sitter recurrence time is exponentially far in the future, but it's still only \emph{finitely far}. Without understanding the ultimate fate of de Sitter space vs. quantum mechanics, there is nothing we can say with any confidence about such timescales. This is not to say that there is some specific problem with our essay's main argument -- we wish simply to express appropriate humility in the face of ignorance. 

Once again, we stress that, even if our particular picture is wrong, even if there's no de Sitter future, one should view the fine-tuning problems of $m_H$ and $\Lambda$ with skepticism. A careful analysis of the underlying assumptions, along with the blurring of the UV$\rightarrow$IR arrow by quantum gravity, should convince us that these aren't problems that \emph{must} be solved by some extension of our (GR + Standard Model) effective bulk field theory. Rather, it is reasonable to expect that they will be resolved by quantum-gravitational means. In the above, we tried to outline the implications of such a scenario.

\section*{Acknowledgements}

I am grateful to Scott Melville, Gerben Venken, LinQing Chen, Sudip Ghosh, Slava Lysov, Henry Stoltenberg and Tomonori Ugajin for discussions. This work was supported by the Quantum Gravity Unit of the Okinawa Institute of Science and Technology Graduate University (OIST).

\end{document}